**Perspective: Towards the predictive discovery of ambipolarly dopable ultra-wide-band-gap semiconductors: the case of rutile GeO$_2$.**

Sieun Chae,[1] Kelsey Mengle,[1] Kyle Bushick,[1] Jihang Lee,[1] Nocona Sanders,[1] Zihao Deng,[1] Zetian Mi,[2] P. F. P. Poudeu,[1] H. Paik,[3] J. T. Heron,[1,†] and E. Kioupakis[1,*]

[1] Department of Materials Science and Engineering, University of Michigan, Ann Arbor, MI 48109, USA.

[2] Department of Electrical and Computer Engineering, University of Michigan, Ann Arbor, MI 48109, USA.

[3] Platform for the Accelerated Realization, Analysis, and Discovery of Interface Materials (PARADIM), Cornell University, Ithaca, New York 14853, USA

**Abstract**

Ultrawide-band-gap (UWBG) semiconductors are promising for fast, compact, and energy-efficient power-electronics devices. Their wider band gaps result in higher breakdown electric fields that enable high-power switching with a lower energy loss. Yet, the leading UWBG semiconductors suffer from intrinsic materials limitations with regards to their doping asymmetry that impedes their adoption in CMOS technology. Improvements in the ambipolar doping of UWBG materials will enable a wider range of applications in power electronics as well as deep-UV optoelectronics. These advances can be accomplished through theoretical insights on the limitations of current UWBG materials coupled with the computational prediction and experimental demonstration of alternative UWBG semiconductor materials with improved doping and transport properties. As an example, we discuss the case of rutile GeO$_2$ (r-GeO$_2$), a water-insoluble GeO$_2$ polytype which is theoretically predicted to combine an ultra-wide gap with ambipolar dopability, high carrier mobilities, and a higher thermal conductivity than β-Ga$_2$O$_3$. The subsequent realization of single-crystalline r-GeO$_2$ thin films by molecular beam epitaxy provides the opportunity to realize r-GeO$_2$ for electronic applications. Future efforts towards the predictive discovery and design of new UWBG semiconductors include advances in first-principles theory and high-performance computing software, as well as the demonstration of controlled doping in high-quality thin films with lower dislocation densities and optimized film properties.

**Introduction**

Modern semiconductor device technology advances with the development of new semiconducting materials. Commercial semiconductors such as Si, Ge, and the III-V families (e.g., nitrides, phosphides, arsenides, etc.) are both n-type and p-type dopable, which enables a wide variety of devices such as visible light-emitting diodes and field-effect transistors. For efficient high-power electrical conversion and UV light emission, however, UWBG semiconductors with gaps wider than the 3.4 eV gap of GaN are necessary. AlN and high-Al-content AlGaN, $\beta$-Ga$_2$O$_3$, diamond, and c-BN have all emerged as candidate materials to advance the frontier in high-power electronics.[1] The Baliga figure of merit (BFOM = $\frac{1}{4}\varepsilon_0 \mu E_c^3$ where $\varepsilon_0$ is the static dielectric constant, $\mu$ is the carrier mobility, and $E_c$ is the dielectric breakdown field) is commonly used to quantify the interplay between the breakdown voltage and energy dissipation through resistive losses, and thus to benchmark the efficiency of materials for power devices. Table I lists the BFOM and thermal conductivity for current state-of-the-art UWBG materials. Due to the cubic dependence of the BFOM on the dielectric breakdown field, the most promising materials are those with band gaps wider than 3.4 eV.

Table I. Baliga's figure of merit (BFOM = $\frac{1}{4}\varepsilon_0\mu E_C^3$) and thermal conductivity for silicon and common ultra-wide-band-gap semiconductors. $\varepsilon_0$ is static dielectric constant, $\mu_e/\mu_h$ is electron/hole mobility at room temperature, $E_C$ is dielectric breakdown field predicted based on the breakdown versus band gap relation established by Higashiwaki et al[2], $E_d/E_a$ is donor/acceptor ionization energy, and $\kappa$ is thermal conductivity at room temperature. $\mu_e/\mu_h$ is experimental maximum realized values for all materials except r-GeO$_2$ whereas $\mu_e/\mu_h$ of r-GeO$_2$ is phonon-limited mobility calculated by density functional theory.

| Material | $\varepsilon_0$ | $\mu_e/\mu_h$ (cm$^2$ V$^{-1}$ s$^{-1}$) | $E_C$ (MV cm$^{-1}$) | $E_d/E_a$ (eV) | n-/p-BFOM (10$^6$ V$^2$ Ω$^{-1}$ cm$^{-2}$) | $\kappa$ (W m$^{-1}$ K$^{-1}$) |
|---|---|---|---|---|---|---|
| Si | 11.9[3] | 1240[4]/450[5] | 0.3[2] | 0.04/0.05[6] | 8.8[7]/3.2 | 130[6] |
| 4H-SiC | 9.7[4] | 980[4]/120[8] | 2.5[2] | 0.05/0.19[9] | 3,300[7]/404 | 370[1] |
| GaN | 10.4 (∥c)[10] | 1000[8]/31[11] | 3.3[2] | 0.04/0.21[12] | 8,300/257[7] | 253[1] |
| β-Ga$_2$O$_3$ | 10.0[13] | 184[14]/- | 6.4[2] | 0.04[15]/1.1[16] | 6,300[7]/- | 11; 27[17] |
| AlN | 9.1[18] | 426[19]/14[20] | 15.4[1] | 0.25[21]/1.4[20] | 336,000[1]/11,000 | 286; 319[22] |
| c-BN | 7.1[23] | 200[23]/500[24] | 17.5[1] | 0.15[25]/0.24[26] | 27,800/695,000[1] | 1600[27] |
| diamond | 5.7[6] | 1060/2000[28] | 13.0[1] | 0.57/0.38[28] | 294,000/554,000[1] | 2290-3450[28] |
| r-GeO$_2$ | 14.5 (⊥c)[29] 12.2 (∥c)[29] | 244/27[7] 377/29[7] | 7.0[7] | <0.04/0.45[30] | 27,000/3,000[7] 35,000/2,700[7] | 51[31] |

The emerging UWBG materials face several doping challenges, however. Specifically, ambipolar doping is a challenge for all current UWBG materials, which limits the application of many UWBG semiconductors to unipolar devices. For Al$_x$Ga$_{1-x}$N, both n-type and p-type doping efficiencies decrease with increasing Al content, $x$, as the ionization energy of the Mg acceptor increases and compensating defects such as N vacancies form more easily with increasing $x$.[32,33] β-Ga$_2$O$_3$ is characterized by flat valence bands that lead to deep ionization energies for acceptors (> 1.1 eV) and the formation of self-trapped hole polarons.[16,34] For c-BN and diamond, n-type doping has proven to be challenging. Due to the small lattice constant of c-BN and diamond, the range of dopants that fit into the lattice is severely limited and the best substitutional donors (currently P for diamond, and S and Si for c-BN) have high activation energies (> 0.4 eV).[35–37]

Additionally, each of the current UWBG materials has its own drawbacks that hamper its adoption in devices. AlGaN and diamond suffer from high synthesis and processing costs, high dislocation densities, and limited size of native substrates. The synthesis of c-BN is also challenging as the hexagonal phase of BN is more stable than the cubic polytype. $\beta$-Ga$_2$O$_3$ is currently the subject of intense research activity due to the availability of affordable semi-insulating native substrates and the ease of n-type doping. However, $\beta$-Ga$_2$O$_3$-based devices are energy-inefficient owing to the relatively low mobility (184 cm$^2$ V$^{-1}$ s$^{-1}$)[14] compared to other UWBG semiconductors, while its poor thermal conductivity (11 W m$^{-1}$ K$^{-1}$ ∥ $\vec{a}$ and 27 W m$^{-1}$ K$^{-1}$ ∥ $\vec{b}$)[17] hinders the removal of the generated heat. Therefore, UWBG semiconductor research must simultaneously seek to improve the performance of current materials in order to realize their full potential, and at the same time to explore novel UWBG materials and critically assess their potential to advance the current state of the art.

In the exploration of new materials, a number of prospective UWBG materials have been proposed. E.g., recent calculations have identified that the rocksalt phase of ZnO is ambipolarly dopable.[38] However, rocksalt ZnO is metastable and has been stabilized only by alloying it with other rocksalt oxides (e.g., NiO or MgO), which sacrifices the mobility and/or doping properties.[39] Spinel ZnGa$_2$O$_4$ has proven to be n-type dopable with a band gap and electron mobility comparable to $\beta$-Ga$_2$O$_3$.[40] Hole conduction, however, has been achieved only at high temperatures (> 600 K)[41] while the thermal conductivity is also low (22.1 W m$^{-1}$ K$^{-1}$)[40]. LiGaO$_2$ is also a potential ultrawide-band-gap (5.8 eV) semiconductor which is theoretically predicted to be n-type dopable with Si or Ge, though experimental investigation is needed to realize its potential application.[42–44]

In our recent work, we identified rutile GeO$_2$ (r-GeO$_2$) as a promising UWBG ($E_g$ = 4.68 eV) semiconductor with ambipolar dopability, high thermal conductivity, and high BFOM (Table I). Further, the stabilization of single-crystalline r-GeO$_2$ bulk crystals and thin films make experimental investigations feasible. In this perspective, we review the key properties of r-GeO$_2$ as an UWBG semiconductor, and we articulate challenges and opportunities for the field.

**R-GeO$_2$ crystal structure, band structure, and effective masses**

In the periodic table, Ge is the group 14 element between Si and Sn, sitting in the fourth period next to Ga. Accordingly, GeO$_2$ has an ultra-wide band gap similar to that of Ga$_2$O$_3$ but adopts chemical and structural properties analogous to SiO$_2$ or SnO$_2$. Though both Ga$_2$O$_3$ and SnO$_2$ are established wide-band-gap n-type semiconductors, little has been known about the semiconducting properties of GeO$_2$ until recently. Among the multiple polymorphs of GeO$_2$, the octahedrally coordinated rutile structure is the high-density crystalline polytype and the thermodynamically most stable phase up to 1030 °C.[45] In contrast to quartz or amorphous GeO$_2$, the rutile phase is insoluble in water,[46] thus it is better suited for device processing.

The crystal structure of rutile $GeO_2$ (r-$GeO_2$) is shown in Figure 1(a). Considering the anisotropy of the rutile crystal structure, its optical or transport properties are often studied along different crystallographic directions (e.g., ⊥c and ∥c). The electronic band structure of r-$GeO_2$ is calculated using different methods such as LDA,[47] GGA,[48] HSE06,[30] and $G_0W_0$.[49,50] The $G_0W_0$-calculated quasiparticle band structure of r-$GeO_2$ is presented in Figure 1(b). The fundamental band gap of r-$GeO_2$ is direct at Γ, with a calculated value (4.44 eV) that is close to experimental UV-absorption measurements (4.68 eV).[51] The direct optical gap between the VBM and CBM is dipole forbidden and the direct allowed transitions occur from VBs approximately 0.6 eV (2.21 eV) below VBM to CBM for ⊥c (∥c) direction (Table III in Ref. 49) This band-gap value falls into the UWBG region, making r-$GeO_2$ promising for power electronics applications.

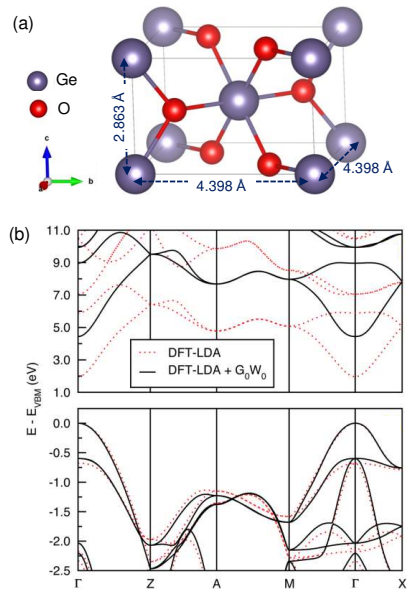

Figure 1. (a) Crystal structure and (b) electronic band structure of r-$GeO_2$ calculated within DFT-LDA (red dotted lines) and with the quasiparticle DFT-LDA + $G_0W_0$ method (black solid lines).

Reproduced from Mengle et al., J. Appl. Phys. **126**, 085703 (2019) with the permission of AIP Publishing.

Another band feature is the band dispersion, i.e., the effective mass ($m^*$). The carrier effective masses are key parameters that determine the n-/p-type dopability, since according to the Bohr model, the shallow donor/acceptor ionization energy ($E_{d/a}$) is proportional to electron/hole effective mass ($E_{d/a} = 13.6 \cdot \frac{m^*_{e/h}}{\varepsilon_r^2}$ eV). The Drude mobility ($\mu = \frac{e \cdot \tau}{m^*}$) and the coupling constant of polarons are also determined by $m^*$,[52] thus lighter effective masses improve the electrical conductivity of semiconductors and suppress the formation of polarons in polar materials. However, the effective mass generally becomes heavier as the band gap increases,[52] which makes most UWBG materials unsuitable for electronic applications.

Despite its ultra-wide band gap, r-GeO$_2$ exhibits relatively light electron and hole effective masses. The effective-mass values of r-GeO$_2$ are obtained by fitting the hyperbolic equation to the G$_0$W$_0$ band structure. The electron effective masses along $\Gamma \to$ X ($m^*_{e\perp}$) and $\Gamma \to$ Z ($m^*_{e\parallel}$) are 0.43 $m_0$ and 0.23 $m_0$, respectively.[49] These values are similar to other n-type semiconductors such as $\beta$-Ga$_2$O$_3$ (0.23 – 0.34 $m_0$)[53], SnO$_2$ (0.23 – 0.30 $m_0$)[54], and GaN (0.19 – 0.21 $m_0$)[55]. In addition, the hole effective masses ($m^*_{h\perp}$=1.28 $m_0$ and $m^*_{h\parallel}$=1.74 $m_0$[49]) are notably small compared to other common ultra-wide-band-gap materials. For $\beta$-Ga$_2$O$_3$, the valence band is notoriously flat,[56] resulting in a large hole effective mass of ~40 $m_0$ which also gives rise to trapped hole polarons with a trapping energy of 0.53 eV.[34] In contrast, in the absence of impurities, the self-trapped energy of hole polaron in r-GeO$_2$ is calculated to be less than 0.01 eV,[49] indicating its superior hole-transport properties compared to $\beta$-Ga$_2$O$_3$.

Why are the effective masses of r-GeO$_2$ lower than materials with similar band gaps? First, the conduction bands consist of delocalized Ge 4$s$ orbitals, leading to a broad conduction band width. Moreover, while the top valence bands consist of localized O 2$p$ orbitals, the densely packed oxygen atoms allow holes to conduct easily through oxygen orbitals. The delocalized nature of electrons and holes makes r-GeO$_2$ promising for ambipolar dopability.

**Donors and Acceptors**

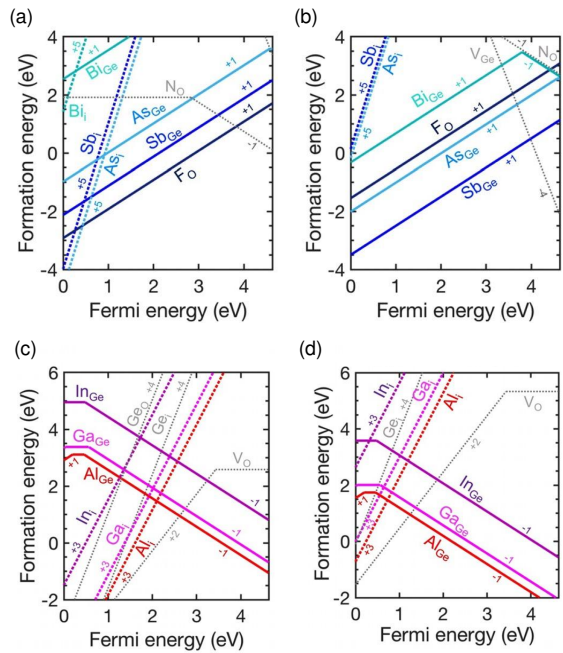

Figure 2. Formation energy of (a-b) donor defects and (c-d) acceptor defects along with potential charge-compensating native defects as a function of the Fermi level. The simulated growth

conditions are (a, c) Ge rich/O poor and (b, d) O rich/Ge poor conditions. Reproduced from Chae et al., Appl. Phys. Lett. **114**, 102104 (2019), with the permission of AIP Publishing.

To identify potential donors and acceptors in r-GeO$_2$, Chae and colleagues have applied hybrid density functional theory to calculate the formation and ionization energies of dopants and to identify possible sources of charge compensation.[30] From the calculations, it is predicted that Sb$_{Ge}$, As$_{Ge}$, and F$_O$ are all shallow donors with an ionization energy of ~25 meV (Figure 2(a-b)). The incorporation of donors varies depending on the growth conditions: F$_O$ is favored under O-poor/Ge-rich growth, while Sb$_{Ge}$ is the donor with the lowest formation energy under O-rich/Ge-poor conditions. Under O-rich/Ge-poor conditions, however, the Ge vacancy is an unavoidable defect that compensates donors. Therefore, O-poor/Ge-rich growth conditions are preferred to enable n-type doping under thermodynamic equilibrium, since the only compensating defects we identified are nitrogen impurities (N$_O$), which however can be eliminated by excluding N from the growth environment.

Group 13 elements such as Al, Ga, and In are possible acceptors in r-GeO$_2$, among which Al$_{Ge}$ is calculated to be the best candidate acceptor due to its low ionization energy and low formation energy (Figure 2(c-d)). While hole polarons do not form in the absence of impurities, all acceptors examined in the study form a hole polaron in the neutral charge state accompanied by local lattice distortions, resulting in a relatively high ionization energy (0.45 eV for Al$_{Ge}$). Also, charge compensation from donor-type native defects such as V$_O$ or self-passivating defects (Al$_i$) can be a challenge for p-doping of r-GeO$_2$. However, due to the strong Coulombic interaction between Al$_{Ge}$ and hydrogen interstitial (H$_i$), co-doping with H$_i$ can effectively enhance the solubility of Al$_{Ge}$ up to the Mott-transition limit (~$10^{20}$ cm$^{-3}$), allowing an impurity band to form and reducing the effective ionization energy. P-type conduction is enabled through the impurity band by post-annealing removal of $H_i$ to activate hole carriers. Thermally activated p-type conduction of r-GeO$_2$ has also been demonstrated experimentally by Niedermeier et al.[57] The notably shallow acceptor ionization of r-GeO$_2$ compared to other established WBG oxide semiconductors (e.g., 0.91 eV for SnO$_2$[58]) originates from the dense network of O atoms in r-GeO$_2$ that allows strong O 2$p$ antibonding interactions and leads to a high-lying valence-band maximum (VBM) (Supplementary Information of Ref.[30], and Ref.[57]) and light hole effective masses. This is in contrast to the valence bands of most oxide semiconductors which are deep (thus inducing high ionization potentials).

**Mobility, BFOM, and thermal conductivity**

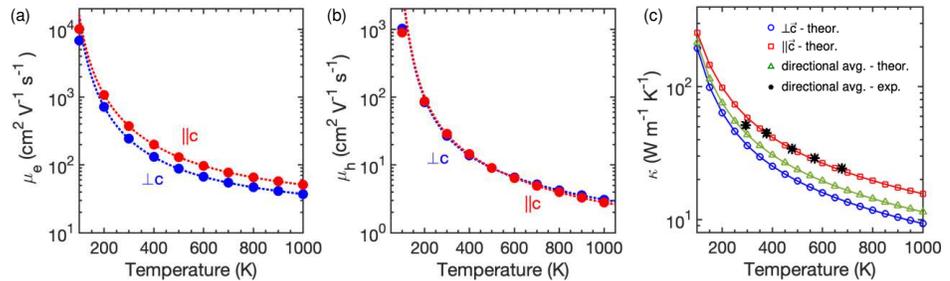

Figure 3. (a) Phonon-limited electron and (b) hole mobility of r-GeO$_2$ along the ⊥c and ∥c directions as a function of temperature for a carrier concentration of n = $10^{17}$ cm$^{-3}$. Data from Ref.[59] (c) The theoretically calculated and experimentally measured thermal conductivity of polycrystalline r-GeO$_2$ from 100 K to 1000 K. Reproduced from Chae et al., Appl. Phys. Lett. **117**, 102106 (2020), with the permission of AIP Publishing.

For energy-efficient power electronic devices, a high carrier mobility ($\mu$) for efficient carrier transport, a high breakdown field ($E_C$) and dielectric constant ($\varepsilon_0$) for high voltage operation, and a high thermal conductivity ($\kappa$) for efficient heat extraction are necessary. Bushick et al.[7] determined the phonon-limited electron and hole mobilities of r-GeO$_2$ as a function of temperature and crystallographic orientation by applying density functional and density functional perturbation theories to calculate carrier-phonon coupling in r-GeO$_2$ (Figure 3(a-b)). At 300 K, the calculated electron mobilities are $\mu_{elec,\perp\vec{c}}$ = 244 cm$^2$ V$^{-1}$ s$^{-1}$ and $\mu_{elec,\parallel\vec{c}}$ = 377 cm$^2$, and the calculated hole mobilities are $\mu_{hole,\perp\vec{c}}$ = 27 cm$^2$ V$^{-1}$ s$^{-1}$ and $\mu_{hole,\parallel\vec{c}}$ = 29 cm$^2$ V$^{-1}$ s$^{-1}$. The polar-optical modes exhibit the strongest carrier-phonon coupling and the low-energy polar-optical modes limit the room-temperature mobility in r-GeO$_2$. The calculated electron mobility is comparable to currently used n-type semiconductors. Also, the hole mobility or r-GeO$_2$ is comparable to that of p-type GaN, again showing its promising properties for p-type conduction.

Further computational and experimental results also point to favorable thermal properties for r-GeO$_2$. Figure 3(c) shows the theoretically predicted and the experimentally measured thermal conductivity of r-GeO$_2$ as a function of temperature.[31] First-principles calculations predict an anisotropic phonon-limited thermal conductivity of 37 W m$^{-1}$ K$^{-1}$ ($\perp$c) and 57 W m$^{-1}$ K$^{-1}$ ($\parallel$c) at 300 K. Experimentally, the thermal conductivity was measured using the laser-flash method for hot-pressed, polycrystalline r-GeO$_2$ pellets with grain sizes of ~1.50 $\mu$m. The measured value for r-GeO$_2$ (51 W m$^{-1}$ K$^{-1}$ at 300 K) is approximately 2 times higher than the highest value of $\beta$-Ga$_2$O$_3$ (11 W m$^{-1}$ K$^{-1}$ and 27 W m$^{-1}$ K$^{-1}$ along the $\vec{a}$ and $\vec{b}$ directions, respectively). Also, while $\beta$-Ga$_2$O$_3$ can be only grown on thermally insulating substrates (e.g., Al$_2$O$_3$), the higher symmetry of r-GeO$_2$ allows it to be epitaxially grown on thermally conductive materials such as $SnO_2$ (100 W m$^{-1}$ K$^{-1}$).[31]

By combining the calculated results for the mobility and dielectric constant, the BFOM of r-$GeO_2$ can be evaluated in Table I by using a breakdown field extracted from the breakdown field versus band gap relation established by Higashiwaki et al.[2] While common ultra-wide-band-gap materials suffer from doping asymmetry, r-$GeO_2$ has relatively low dopant ionization energies for both donors and acceptors, and in combination with a higher thermal conductivity and a higher BFOM compared to $\beta$-$Ga_2O_3$, the results demonstrate the promise of r-$GeO_2$ to advance the state of the art in UWBG semiconductor device technology.

**Towards experimental demonstration**

To date, experimental reports on r-$GeO_2$ are largely focused on its synthesis.[60–65] Here we summarize the current advances and challenges of r-$GeO_2$ synthesis. The rutile polymorph of $GeO_2$, with $Ge^{4+}$ ions in the octahedral coordination, is the thermodynamically stable phase at ambient pressure and temperature (Figure 4(a)). R-$GeO_2$ transforms to the quartz phase ($Ge^{4+}$ in tetrahedral coordination) above 1000 ºC and before melting.[46] Unlike $SiO_2$ and $SnO_2$, however, which are stable in the quartz and rutile structure respectively, both the quartz and the rutile are deeply stable polymorphs of $GeO_2$ under ambient conditions.[45] Similarly to $SiO_2$, $GeO_2$ is also a glass former with a deeply metastable amorphous phase. Thus, one of the challenges in the synthesis of r-$GeO_2$ is navigating the kinetic and thermodynamic space to avoid the formation of the deleterious metastable quartz and amorphous phases.

The solid-state synthesis of r-GeO$_2$ can be achieved from commercially available quartz phase powder. As the phase transformation from quartz to rutile is accompanied by a large volume reduction of ~50% and must overcome a large energy barrier of ~400 kJ/mol, the phase transformation occurs with a pressure higher than 100 MPa and a temperature higher than 900 K (~0.45 T$_m$),[31,60,66] which can be achieved by hot pressing. Figure 4(b) shows x-ray diffraction data of the quartz-GeO$_2$ power precursor and subsequent r-GeO$_2$ pellets converted in a hot press.[31] The grain sizes of hot-pressed r-GeO$_2$ pellets range from 0.5 – 2 $\mu$m (Figure 4(c)). At T < 600K and P > 6 GPa conditions, a displacive transition occurs which changes the coordination of the Ge atom from four-fold to six-fold. The resulting phase is amorphous or distorted rutile, depending on the starting materials and pressing conditions.[61–63] While the synthesis of r-GeO$_2$ has been demonstrated through phase conversion, the small grain sizes are undesirable for modern power electronics as grain boundaries or voids act as charge-trapping or scattering centers and degrade device performance.

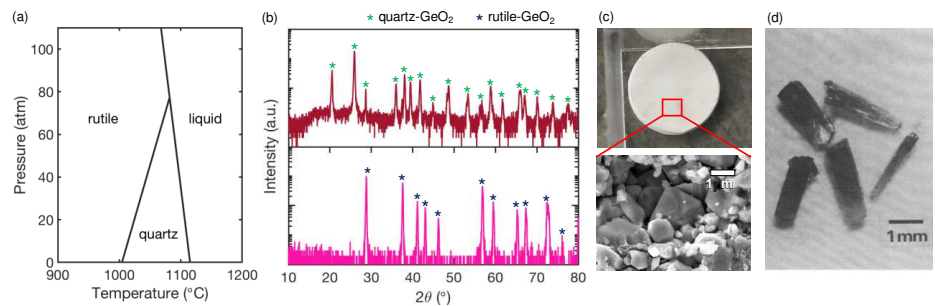

Figure 4. (a) Phase diagram of GeO$_2$. The rutile is the most stable polytype under ambient conditions. (b) The x-ray diffraction of (top) GeO$_2$ powder and (bottom) a GeO$_2$ pellet after hot pressing (800 ºC, 100 MPa) and subsequent annealing (1000 ºC, air). (c) A digital image and scanning electron microscope image of a hot-pressed GeO$_2$ pellet. (d) Single crystals of r-GeO$_2$ synthesized by chemical vapor transport. Panel (a) is adapted with permission from Hill et al. American Mineralogist **53**, 1744 (1968). Copyright 1968 Mineralogical Society of America. Panels (b-c) are reproduced from Chae et al. Appl. Phys. Lett. **117**, 102106 (2020), with the permission of AIP Publishing. Panel (d) is adapted with permission from Agafonov et al. Materials Research Bulletin, **19**, 233 (1984). Copyright 1984 Elsevier B. V.

Various bulk synthesis techniques have been attempted to realize r-GeO$_2$ single crystals. The melting temperature of GeO$_2$ is relatively low (1100 ºC) and conventional crystal growth techniques from the melt (e.g., Czochralski or float zone) have successfully stabilized r-GeO$_2$. However, since quartz is the high-temperature stable solid phase, the solid-state phase change to the lower-temperature r-GeO$_2$ leads to significant internal cracking.[65] Instead, Goodrum[65] utilized alkali-oxide solvents to lower the liquidus temperature below the rutile-to-quartz transition temperature and reported the growth of 10 mm-long r-GeO$_2$ single crystals using the top-seeded flux technique. Single-crystal r-GeO$_2$ growth by chemical vapor transport is also reported (Figure 4(d)).[64] Owing to the high vapor pressure of GeO$_2$, GeO molecules can easily desorb above 700 ºC and the re-condensation of these molecules has been successful for stabilizing the rutile phase. Agafonov et al.[64] utilized TeCl$_4$ and HCl as transport agents to carry GeO molecules and synthesized a single-crystal r-GeO$_2$ rod with a size of 0.5 × 0.5 × 2 mm$^3$ using a temperature gradient of 1000 - 900 ºC. These bulk synthesis studies have demonstrated the possibility of stabilizing bulk single crystals of GeO$_2$ in the rutile polymorph despite the existence of the competing quartz phase near the melting temperature. However, more studies are required to obtain highly crystalline large single crystals of r-GeO$_2$ that can be potentially used to produce substrates for the homoepitaxy of r-GeO$_2$ thin films aimed for electronic applications.

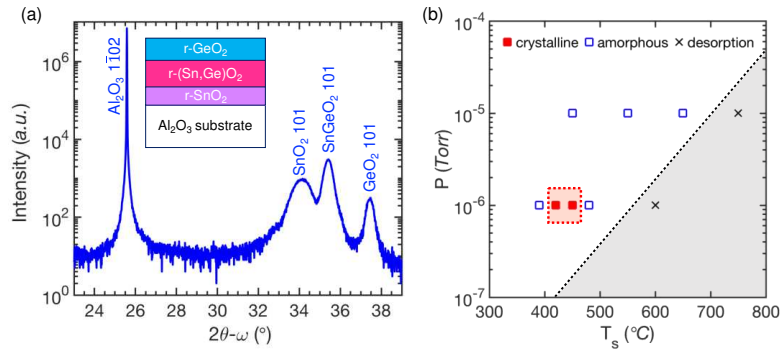

Figure 5. (a) The x-ray diffraction of r-GeO$_2$ thin films grown on (Sn,Ge)O$_2$/SnO$_2$-buffered sapphire substrates using molecular beam epitaxy. (b) The substrate temperature ($T_s$) and pressure ($P$) map for GeO$_2$ deposition and the resulting crystallinity of the GeO$_2$ films. Reproduced from Chae et al. Appl. Phys. Lett. **117**, 072105 (2020), with the permission of AIP Publishing.

The thin-film growth of r-GeO$_2$ is challenging owing to the presence of the deeply metastable glass phase and the high vapor pressure of GeO$_2$. Prior works report the growth of GeO$_2$ films using sputtering,[67–70] pulse laser deposition,[71–73] and thermal evaporation,[74] but the as-deposited films are all amorphous, indicating a strong tendency for glass formation. Recently, however, epitaxial single-crystalline thin films of r-GeO$_2$ were successfully synthesized using ozone-assisted molecular beam epitaxy (MBE)[75] on R-plane sapphire. The R-plane sapphire is a suitable substrate for rutile oxide thin films due to the rectangular surface symmetry and the edge-sharing connectivity of the oxygen octahedra. Figure 5(a) shows the x-ray diffraction of r-GeO$_2$ thin films grown on (Sn,Ge)O$_2$/SnO$_2$ buffered sapphire substrates. In agreement with prior SnO$_2$ thin-film growth[76–78], the in-plane registry is [010] GeO$_2$ ∥ [11$\bar{2}$0] Al$_2$O$_3$ and [$\bar{1}$01] GeO$_2$ ∥ [$\bar{1}$101] Al$_2$O$_3$. It was found that the stabilization of single-crystalline rutile films requires a balance of epitaxial strain, adatom mobility, GeO desorption, and chemical composition. A compositional dependence of the (Sn,Ge)O$_2$ buffer layer on the stabilization of the r-GeO$_2$ film shows that a strain of 4.4 % ∥ [010] and 5.0 % ∥ [$\bar{1}$01] yields single-crystalline films, whereas a strain of 4.8 % ∥ [010] and 5.8 % ∥ [$\bar{1}$01] causes amorphization. Further, the substrate temperature and ozone pressure must be optimized to allow sufficient adatom mobility without high desorption as well as to achieve proper stoichiometry. Single-crystalline films were realized in a narrow region of growth-parameter space using a preoxidized molecular source to balance the stoichiometry of the films (Figure 5(b)). The amorphous phase emerges when the adatom mobility is insufficient to enable crystal growth or if the film composition deviates too far from stoichiometry. Meanwhile, high temperatures cause GeO desorption, preventing film growth. The synthesis of single-crystalline r-GeO$_2$ thin films provides opportunities to experimentally validate its properties for UWBG semiconductor applications.

**Perspective on theory**

First-principles materials calculations based on density functional theory and related techniques were instrumental in identifying the desirable dopant and transport features of r-GeO$_2$ for power-electronics applications. Such atomistic computational methods, that are made available to the research community through well-maintained open-source computer software, have achieved a sufficient level of predictive accuracy to stimulate and guide experimental synthesis and characterization efforts for the discovery of new UWBG semiconductors that advance the current state of the art. For example, the GW method as implemented in, e.g., the BerkeleyGW software[79] predicts accurate band structures, band gaps, effective masses, and dielectric properties of bulk materials and nanostructures. Moreover, the accurate calculations of electron-phonon coupling properties with density functional perturbation theory in the Quantum ESPRESSO suite of codes and their efficient interpolation for arbitrary wave vectors in the Brillouin zone using maximally localized Wannier functions and the EPW code have enabled accurate and efficient calculations of carrier mobilities through the iterative solution of the Boltzmann transport equation.[80,81] The Boltzmann transport equation applied for phonons with software such as almaBTE can also predict the phonon-limited thermal conductivity of materials.[82] Hybrid-functional calculations have also been instrumental in predicting defect formation energies and charge-transition levels in a wide range of semiconductors. Future advanced in computational method and software development will be instrumental in advancing the predictive theoretical characterization of new UWBG semiconductors. E.g., techniques to understand carrier scattering by phonons[83], defects[84], and alloy disorder[85], as well as dielectric breakdown phenomena under intense electric fields[86], as well as the formation of defects and two-dimensional electron/hole gases at semiconductor interfaces[87,88] will be important.

Moreover, first-principles calculations in combination with modern high-performance computing recourses, automated high-throughput calculation execution, open-access materials databases, and materials-informatics techniques can accelerate the discovery of advanced semiconductors through the identification of structural and chemical materials features that give rise to desirable functionalities. E.g., the material features that make r-$GeO_2$ a promising semiconductor for power electronics are (1) its highly symmetric crystal structure, which reduces the modes for electron and phonon scattering and enables high thermal conductivity and mobility, and (2) its dense packing of oxygen anions, which produces to a strong overlap of the O 2$p$ orbitals and consequently a small hole effective mass that avoids polaron formation, reduces the acceptor activation energy, and enables p-type conductivity. Performing calculations in a systematic fashion can identify more materials with unexpected properties. Similar high-throughput calculations have been successfully applied to discover new p-type transparent conducting materials[89–91] using the band gap and hole effective mass as the screening parameters to predict the competing properties of optical transparency and electrical conductivity. Subsequently, higher-accuracy calculations are performed for the most promising candidates to confirm their desirable band structure, mobility, and dopability. In the field of UWBG semiconductors, Gorai et al.[92] performed a broad computational survey to identify materials with a high Baliga figure of merit and high lattice thermal conductivity using materials-informatics relations (derived from earlier high-throughput calculations) that link the transport properties (mobility, thermal conductivity) and critical breakdown field to intrinsic material parameters such as dielectric constant, effective mass, phonon cutoff frequency and bulk modulus. Similar analyses can be deployed to identify candidate UWBG materials with shallow dopants and mobile carriers that can be tested experimentally.

**Perspectives on synthesis**

On the experimental side, single-crystalline r-GeO$_2$ thin films have been grown on R-plane sapphire substrates, however, due to the different crystal structure and large lattice mismatch, defects and dislocations are unavoidably introduced which degrade carrier mobility and act as a passivation source for free carriers. Homoepitaxy has many advantages for obtaining high-quality films and achieving better device performance. Bulk single crystals of r-GeO$_2$ have been synthesized through flux or chemical vapor transport techniques, however, there are more research opportunities to improve the crystal-quality and size of r-GeO$_2$ crystals. From the bulk phase diagram (Figure 4a), the quartz phase solidifies from the melt under ambient pressure which then presents a challenge to realize large r-GeO$_2$ single crystals using synthesis techniques such as Czochralski (CZ) or float-zone (FZ). The phase stability can be skewed toward the denser rutile phase with applied pressure during the synthesis which may then allow for the rutile phase to solidify directly from the melt. Prior work on the hydrothermal synthesis of r-GeO$_2$ has reported a P-T diagram showing the rutile phase can be synthesized directly from melt at ~7.6 MPa.[93] Thus, high pressure FZ or CZ furnaces, such as those at the NSF PARADIM center where up to 30 MPa can be achieved in optical FZ furnaces, may be suitable to stabilize bulk r-GeO$_2$ single crystals.[94] Large and high-quality crystalline r-GeO$_2$ substrates would open new doors for the investigation of bulk and thin film electronic properties.

**Perspectives on doping**

Controlled doping is one of the most important achievements in semiconductor research. Regarding UWBG semiconductors, the ability to dope is what makes them distinct from insulators and opens device possibilities using junctions, however, as the band gap of material increases doping becomes more challenging. In the case of wide-band-gap nitrides and oxides such as AlGaN or $\beta$-Ga$_2$O$_3$, compensating anion vacancies form easily under p-type doping conditions, resulting in degradation of doping efficiency[32,95], while the high-lying CBM level of diamond and BN leads to large donor ionization energies[28]. The n- and p-type dopability of r-GeO$_2$ has only been theoretically predicted so far, and more efforts are required toward the experimental realization of doping in r-GeO$_2$. The issues with the doping of r-GeO$_2$ presented by theory are (1) a relatively deep acceptor level (0.45 eV), which arises from the relatively low-lying valence bands, (2) the formation of compensating defects (e.g., V$_O$) that are unavoidable under p-type doping conditions, and (3) dislocations or unintentional impurities that are incorporated during synthesis and serve as carrier-trapping centers. However, the doping issues in UWBG semiconductors may be solved by various defect engineering techniques such as co-doping or non-equilibrium growth/processes.

Successful examples of co-doping techniques to improve doping efficiency include deuteration of boron-doped diamond and hydrogen co-doping of Mg-GaN. The defect complex of H-B-H in diamond has proven to be a more efficient donor compared to single phosphorus defects at room temperature.[96] The enhanced n-doping efficiency is explained by the shift of donor level to shallower values for the defect complex. Hydrogen co-doping also allows enhanced dopant solubility while suppressing the formation of compensating defects.[97] Due to its amphoteric propensity, the hydrogen interstitial (H$_i$) easily forms charge-neutral complexes with shallow dopants, while the Coulomb interaction between H$_i$ and dopant ions reduces the formation energy of the complexes, allowing increased dopant solubility. On the other hand, hydrogen is a fast diffuser and can therefore be effectively removed via post-annealing treatment in a hydrogen-poor environment to reactivate free carriers. Improved p-type doping efficiency through hydrogen co-doping has been demonstrated for Mg doped GaN and is the key for the fabrication of InGaN

LEDs (2014 Nobel Prize in Physics).[98,99] A similar defect-engineering technique can be utilized for the p-doping r-GeO$_2$ as the incorporation of hydrogen can effectively enhance the solubility of Al acceptors up to ~10$^{20}$ cm$^{-3}$, while the calculated dissociation energy of Al-H$_i$ (0.96 eV) complexes can be reached by thermal annealing at 700 ºC[30].

Fermi-level engineering is another method to improve the doping efficiency of r-GeO$_2$. Under equilibrium growth of doped samples, the position of the Fermi level lies close to the band edge, which makes the formation energy of undesirable compensating defects comparable or even lower than the dopants. One solution is to utilize non-equilibrium growth conditions and shift the Fermi level during synthesis away from the band edge. This can suppress the dopant-defect compensation and enhance the doping efficiency. Such ideas have proven very effective in both p- and n-doping of Al-rich AlGaN alloys.[100,101] In p-type AlGaN, doping with Mg acceptors is severely hampered by the low formation energy of compensating V$_N$ in conventional epitaxial growth, during which the Fermi level lies near the VBM. However, when the sample is grown at slightly metal (Ga)-rich conditions, a Schottky junction forms at the growth front between the thin Ga metal layer and the AlGaN semiconductor layer during epitaxy, which pins the Fermi level near the middle of the band gap. As a result, the formation energy of Mg acceptors is significantly reduced while the formation energy of compensating V$_N$ increases dramatically. This non-equilibrium junction-assisted epitaxy demonstrates a high hole concentration of ~ 4.5×10$^{17}$ cm$^{-3}$ in Al$_{0.9}$Ga$_{0.1}$N.[100] Similarly, Fermi level control can also be achieved by growing Si-doped AlGaN under above-gap UV illumination. The UV light excites electron-hole pairs and increases the minority hole concentration during the synthesis of n-type AlGaN, which shifts the Fermi level away from the CBM and suppresses the formation of compensating acceptors. This leads to an order of magnitude enhancement in free electron concentration and an improvement of mobility by a factor of 3.[101] These studies demonstrate the promise of creating favorable growth environments for doping with non-equilibrium synthesis techniques, which can also be applied to enhance the doping efficiency of r-GeO$_2$.

**Conclusion**

The discovery of new materials with enhanced material properties through the synergy of computational and experimental approaches can address the doping limitations of current UWBG semiconductors for power-electronic applications. Here we have reviewed the theoretical prediction and experimental synthesis of r-GeO$_2$, an emerging UWBG semiconductor with ambipolar doping, high carrier mobilities, and a higher thermal conductivity than β-Ga$_2$O$_3$. The key material parameter that makes r-GeO$_2$ suitable for energy-efficient power electronics is its highly symmetric and dense crystal structure that induces strong orbital overlaps, which in turn enable small electron and hole effective masses as well as high-lying valence bands that facilitate p-type doping. These features can serve as materials-design principles to computationally discover previously unexplored ambipolarly dopable UWBG semiconductors. We have also discussed the state of the field in terms of the synthesis and characterization of single-crystalline r-GeO$_2$ in bulk and thin-film form and highlight that improved control over defects and dopants is necessary to experimentally realize efficient doping and enable r-GeO$_2$-based electronics.

AUTHOR INFORMATION

Corresponding Authors

†Email: jtheron@umich.edu

*Email: kioup@umich.edu

**Acknowledgements**

The theoretical work was supported by the National Science Foundation through the Designing Materials to Revolutionize and Engineer our Future (DMREF) Program under Award No. 1534221 (band structure, doping, and defect theory), and by the Computational Materials Sciences Program funded by the U.S. Department of Energy, Office of Science, Basic Energy Sciences, under Award No. DE-SC0020129 (phonons, mobility, and thermal-transport theory). The experimental work was supported by the National Science Foundation Award No. DMR 1810119 (bulk and thin film synthesis, structural characterization), by the U.S. Department of Energy, Office of Science, Basic Energy Sciences under Award No. DE-SC00018941 (bulk thermal transport measurements by the National Science Foundation [Platform for the Accelerated Realization, Analysis, and Discovery of Interface Materials (PARADIM)] under Cooperative Agreement No. DMR-1539918 (thin-film synthesis and characterization), and by the NSF CAREER grant DMR-1847847 (perspective on synthesis).

**Data Availability Statement**

Data sharing is not applicable to this article as no new data were created or analyzed in this study.

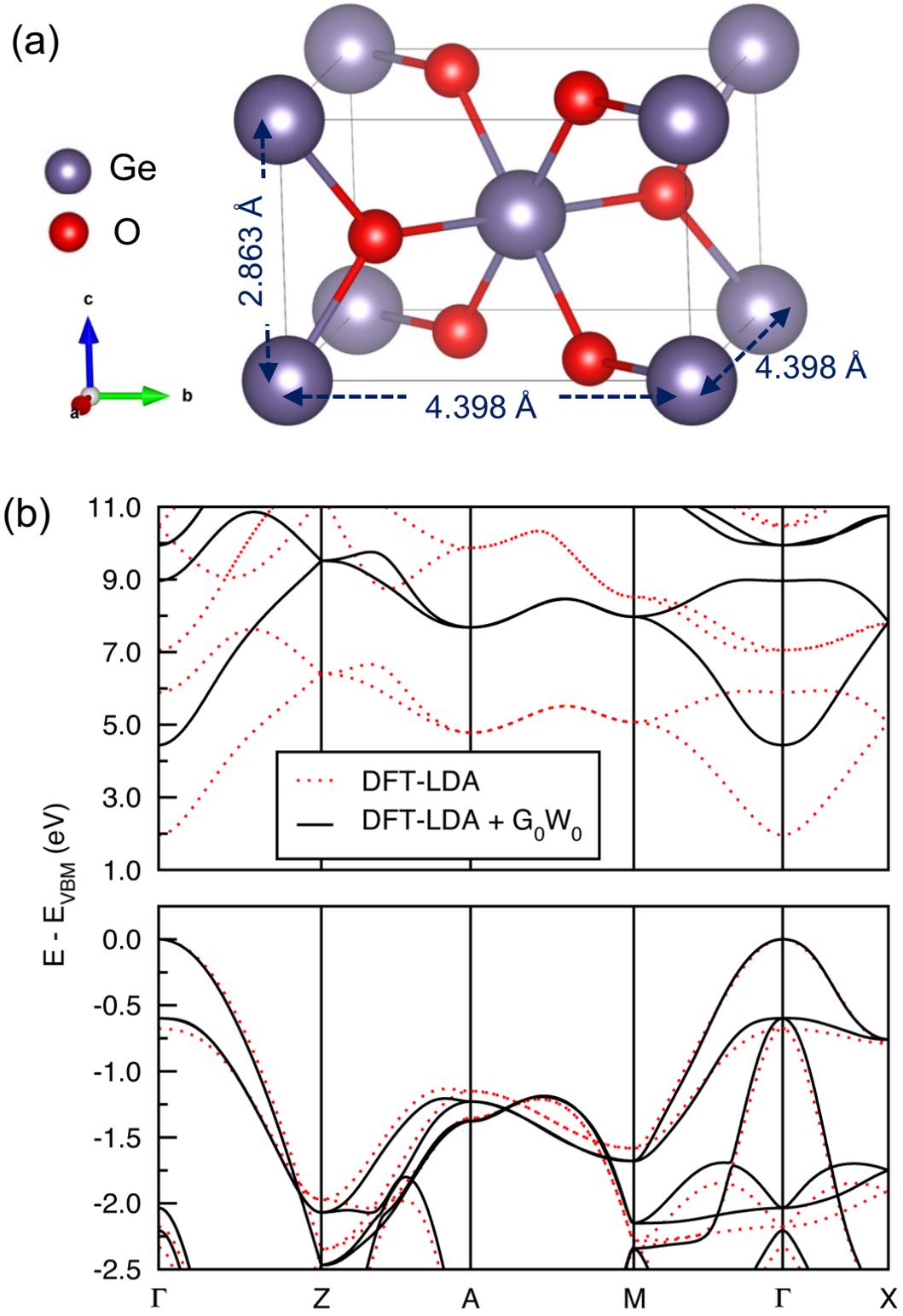

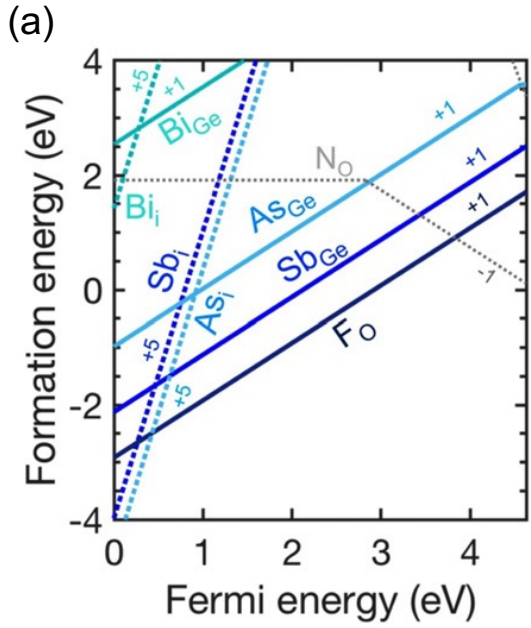
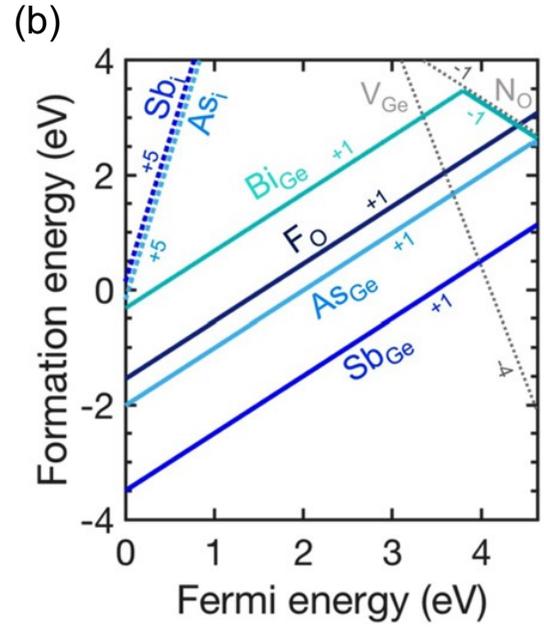
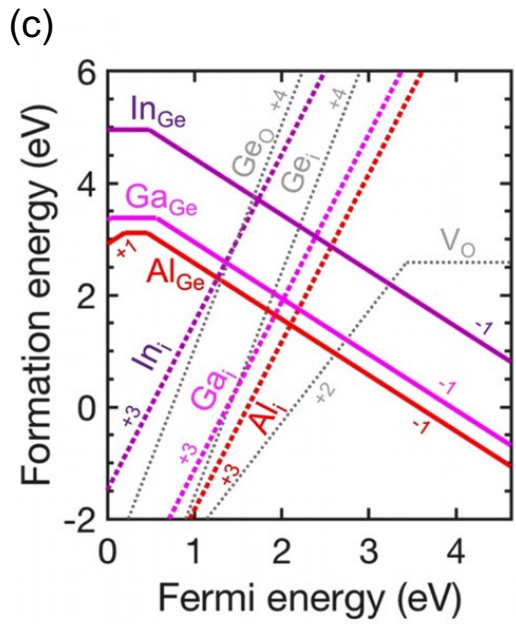
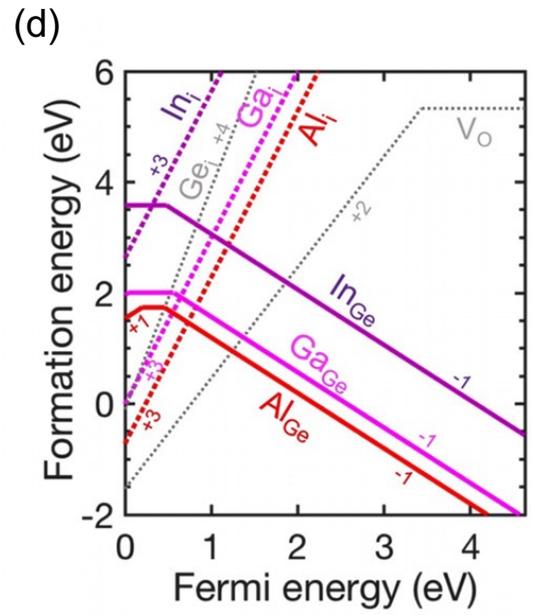

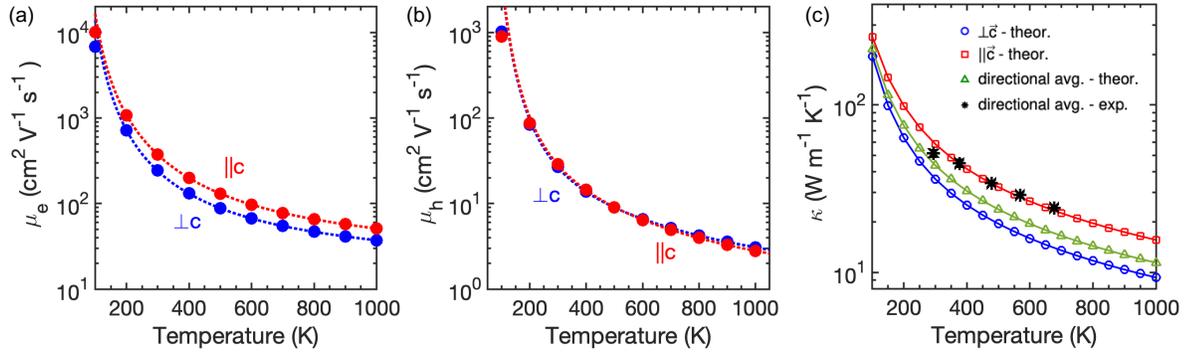

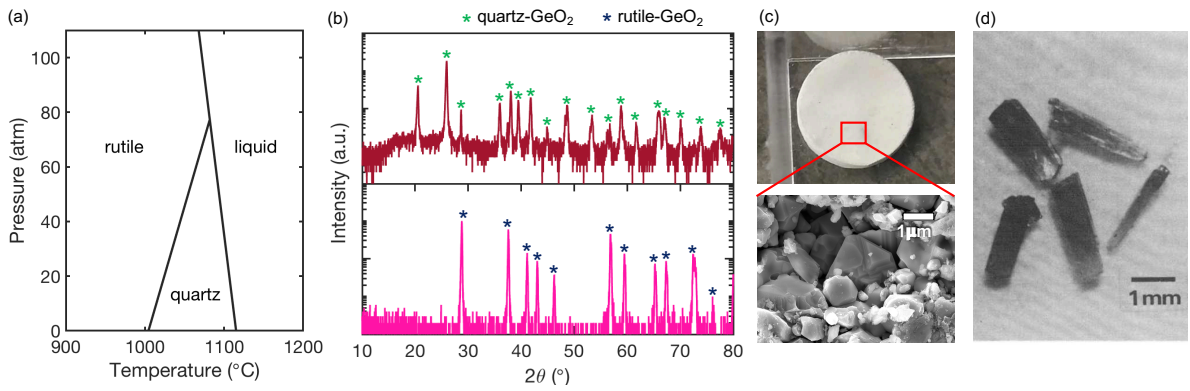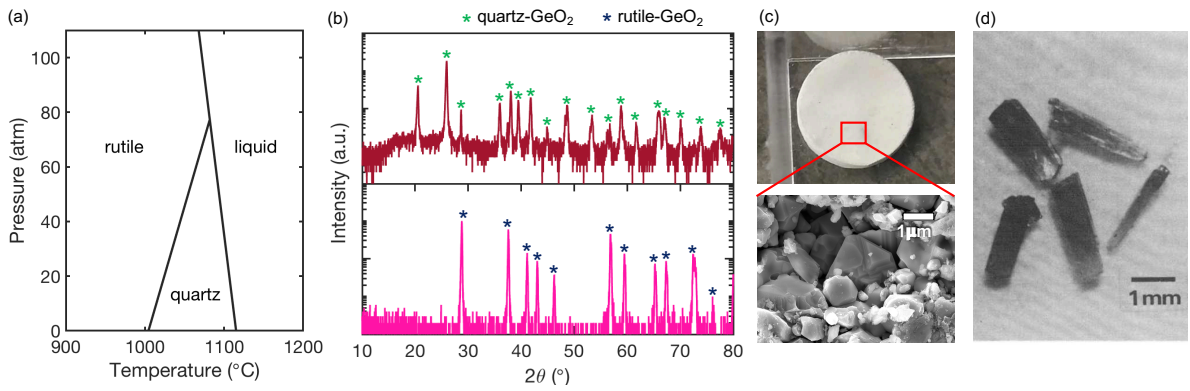

(a) 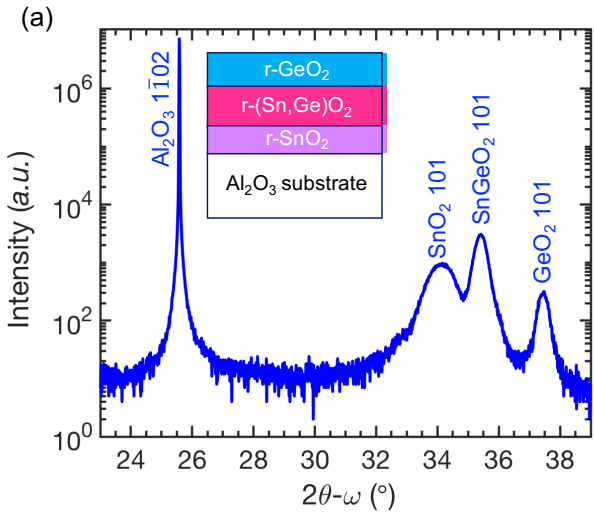
(b) 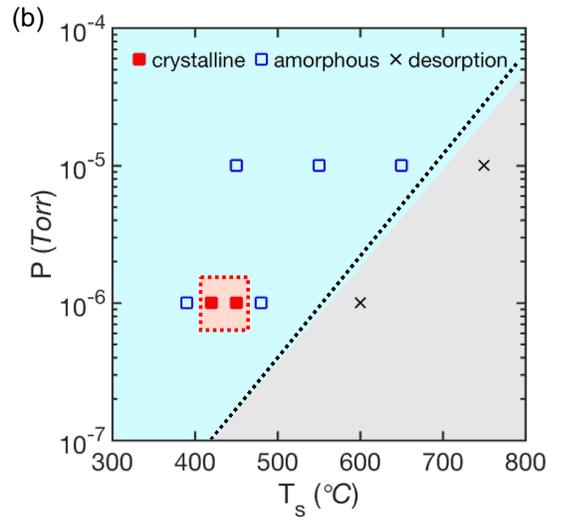